\documentclass[12pt]{article}
\pdfoutput=1 % if your are submitting a pdflatex (i.e. if you have  images in pdf, png or jpg format)
\usepackage{amssymb, amsmath,amsfonts}
\usepackage{graphicx}
\usepackage{xcolor}
\usepackage[pdftex, bookmarks=true,colorlinks,linkcolor=red,urlcolor=blue,citecolor=blue]{hyperref}
\usepackage[normalem]{ulem}
\usepackage{cite}

\def\arXiv#1{\href{http://arxiv.org/abs/#1}{arXiv:#1}}
\def\arXiv#1#2{\href{http://arxiv.org/abs/#1}{arXiv:#1}}

\def\doi#1#2{\href{http://doi.org/#1}{#2}}
\def\blue#1{\textcolor{blue}{#1}}

\makeatletter\@addtoreset{equation}{section}\makeatother

\newcommand{\preprint}[1]{\begin{table}[t]  %%
             \begin{flushright}               %%
             {#1}                             %%
             \end{flushright}                 %%
             \end{table}}                     %%
\renewcommand{\title}[1]{\vbox{\center\LARGE{#1}}\vspace{5mm}}
\renewcommand{\author}[1]{\vbox{\center#1}\vspace{5mm}}
\newcommand{\address}[1]{\vbox{\center\em#1}}

\usepackage{bm}
\def\be{\begin{eqnarray}}
\def\ee{\end{eqnarray}}
\def\bea{\begin{eqnarray}}
\def\eea{\end{eqnarray}}
\newcommand{\nn}{\nonumber}

\def\Dslash{\,\,{\raise.15ex\hbox{/}\mkern-12mu D}}
\def\Dbarslash{\,\,{\raise.15ex\hbox{/}\mkern-12mu {\bar D}}}
\def\delslash{\,\,{\raise.15ex\hbox{/}\mkern-9mu \partial}}
\def\delbarslash{\,\,{\raise.15ex\hbox{/}\mkern-9mu {\bar\partial}}}
\def\pslash{\,\,{\raise.15ex\hbox{/}\mkern-9mu p}}
\def\calDslash{\,\,{\raise.15ex\hbox{/}\mkern-12mu {\cal D}}}

\def\lae{\mathrel{\mathop{\smash{\lower .5 ex \hbox{$\stackrel<\sim$}}}}}
\def\lae{\mathrel{\mathop{\smash{\lower .5 ex \hbox{$\stackrel>\sim$}}}}}

\begin{document}

\unitlength = .8mm

\begin{titlepage}
\vspace{.5cm}
\preprint{}

\begin{center}
\hfill \\
\hfill \\
\vskip 1cm

\title{\boldmath Chiral vortical conductivity across a topological phase transition
from holography}
\vskip 0.5cm
{Xuanting Ji$^{a}$}\footnote{Email: {\tt jixuanting14@mail.ucas.ac.cn}},
{Yan Liu$^{b,c}$}\footnote{Email: {\tt yanliu@buaa.edu.cn}}
and
{Xin-Meng Wu$^{b,c}$}\footnote{Email: {\tt wu\_xm@buaa.edu.cn}}
\address{${}^a$School of Physical Sciences, University of Chinese Academy of Sciences, \\Beijing 100049,  China}
\address{${}^b$Center for Gravitational Physics, Department of Space Science,\\
Beihang University, Beijing 100191, China}
\address{${}^c$Key Laboratory of Space Environment Monitoring and Information Processing,\\ 
Ministry of Industry and Information Technology, Beijing, China}
\end{center}
\vskip 1.5cm

\abstract{ We study the chiral vortical conductivity in a holographic Weyl semimetal model which describes a topological phase transition from the strongly coupled topologically  nontrivial phase to a trivial phase. We focus on the temperature dependence of the chiral vortical conductivity where the mixed gauge-gravitational anomaly plays a crucial role.
After a proper renormalization of the chiral vortical conductivity by the anomalous Hall conductivity and temperature squared,  we find that at low temperature in both the Weyl semimetal phase and the quantum critical region this renormalized ratio stays as universal constants. Furthermore, we provide numerical evidence that this ratio in the quantum critical region depends only on the emergent Lifshitz scaling exponent at the quantum critical point.}

\vfill

\end{titlepage}

\begingroup
\hypersetup{linkcolor=black}
\tableofcontents
\endgroup

%%%%%%%%%%%%%
\section{Introduction}
%%%%%%%%%%%%%

Quantum anomaly induced macroscopic transport physics has attracted great attention recently. On the one hand, the anomalous transports can be studied from many different theoretical approaches, including quantum field theory, chiral kinetic theory, hydrodynamical method and also holographic methods etc. Most interestingly,  universal behaviors for these anomalous transports are found to be completely determined by the anomaly coefficients, including chiral magnetic effect and chiral vortical effect  \cite{Kharzeev:2013ffa, Landsteiner:2016led}. On the other hand, the anomalous transports play a crucial role in the dynamics of many different real physical systems involving chiral fermions, ranging from the quark gluon plasma created at RHIC and LHC \cite{Kharzeev:2015znc}, Neutron stars \cite{Kaminski:2014jda, Shaverin:2014xya}, to the  more recently discovered Weyl semimetals \cite{Landsteiner:2016led, Sun:2016gpy}.

Mixed gauge-gravitational anomaly is one of the novel quantum anomalies, which is the gravitational contribution to the axial anomaly. At finite temperature, the mixed gauge-gravitational anomaly contributes to the chiral vortical conductivity \cite{Landsteiner:2011cp}. More precisely, for a chiral many-body system at finite temperature, the axial current is generated along the direction of rotation in the form of ${\mathbf{J}}_5=\sigma_V {\boldsymbol{\omega}}$ with $\sigma_V=c_g T^2$  and $c_g = 32 \pi^2 \zeta$ where $\zeta$ is the coefficient of the gravitational contribution to
the axial anomaly $\partial_\mu J_5^\mu = \zeta \epsilon^{\mu\nu\rho\sigma}  R^\alpha\,_{\beta\mu\nu} R^\beta\,_{\alpha\rho\sigma} $ \cite{Landsteiner:2011cp, Landsteiner:2011iq}. An incomplete list on the developments of the studies on chiral vortical conductivity is \cite{Erdmenger:2008rm, Banerjee:2008th, Son:2009tf, Neiman:2010zi,Golkar:2012kb,Hou:2012xg, Jensen:2012kj, Basar:2013qia, Jensen:2013rga, Stephanov:2015roa, Golkar:2015oxw, Chowdhury:2016cmh, Glorioso:2017lcn}. It was reported recently in \cite{Gooth:2017mbd} that the effect of positive longitudinal magneto-thermoelectric conductivity linked to mixed gauge-gravitational anomaly has been observed in the Weyl semimetal NbP.

Most of the previous studies of chiral vortical conductivity   focused on systems with time reversal symmetry. However, in Weyl semimetals where low energy excitations behave as chiral fermions, the time reversal symmetry could be explicitly broken
 by the separation of the Weyl nodes and there might be novel effects on chiral vortical conductivity.
Furthermore, Weyl semimetals can go through a topological phase transition to a topological trivial phase. In the critical regime and the trivial phase, since there is no obvious notion of chiral symmetry from the weakly coupled field theory the computation of chiral vortical conductivity as well as other anomalous transports, might be a challenging work. Nevertheless, it is still expected that the chiral vortical conductivity shows  some interesting behaviors across the phase transition.

The calculation of anomalous transports might be easier for strongly coupled Weyl semimetal.
A candidate for strongly coupled Weyl semimetals in the laborary has been recently reported in \cite{Gooth2}.
The goal of this work is to investigate the chiral vortical effect induced from the mixed gauge-gravitational anomaly in strongly interacting Weyl semimetals. The strongly coupled Weyl semimetal can be studied within the framework of holography  \cite{Landsteiner:2015pdh,Landsteiner:2015lsa}, in which the Ward identities of the dual system is the same as the one from weakly coupled field theory, although there is no notion of band structure.
%The goal of this work is to investigate the chiral vortical effect in strongly interacting Weyl semimetals \red{due to} \blue{induced from} the mixed gauge-gravitational anomaly. A strongly interacting Weyl semimetal model has been developed in \cite{Landsteiner:2015pdh,Landsteiner:2015lsa} from the perspectives of gauge/gravity duality in which
It was shown in \cite{Landsteiner:2015pdh,Landsteiner:2015lsa} that in the holographic Weyl semimetal model there is a quantum phase transition between the topologically nontrivial phase and a trivial phase by tuning the ratio between the mass parameter and the time reversal symmetry breaking parameter. Other aspects of the holographic Weyl semimetal have been explored in \cite{Ammon:2016mwa,Landsteiner:2016stv,Grignani:2016wyz,Copetti:2016ewq,Ammon:2018wzb,Baggioli:2018afg,Liu:2018djq,Liu:2018spp}. A recent review on the holographic Weyl semimetals can be found in \cite{review}.

In the following we shall use Kubo formulae to calculate the chiral vortical conductivity in  different phases of holographic Weyl semimetal.
Due to the fact that the axial gauge field is screened in the background of a scalar field, the anomalous Hall conductivity and chiral vortical conductivity should have the same normalization factor.\footnote{More details on this point will be discussed in Sec. \ref{subsec3.1}.} For this reason we use the anomalous Hall conductivity to normalize the chiral vortical conductivity, and study the universality of this renormalised quantity in different phases. We also comment on the behavior of odd (Hall) viscosity in this paper.

This paper is organized as follows. We first review the holographic Weyl semimetal model to fix the conventions in Sec. \ref{sec2}.
In Sec. \ref{sec3}, we calculate the chiral vortical conductivity and study its universality upon normalization by the anomalous Hall conductivity.  We conclude with discussion in Sec. \ref{sec4}. The details of the calculation are in the appendix.

%%%%%%%%%%%%%
\section{Setup}
\label{sec2}
%%%%%%%%%%%%%

We start from the holographic Weyl semimetal model \cite{Landsteiner:2015pdh,Landsteiner:2015lsa}\footnote{Note that
$\epsilon_{abcde}=\sqrt{-g}\varepsilon_{abcde}$ with
$\varepsilon_{0123r}=1$.
}
which is described by the following action
\begin{align}\
  S=&\int d^5x \sqrt{-g}\bigg[\frac{1}{2\kappa^2}\Big(R+\frac{12}{L^2}\Big)-\frac{1}{4e^2}\mathcal{F}^2-\frac{1}{4e^2}F^2
+\epsilon^{abcde}A_a\bigg(\frac{\alpha}{3} \Big(F_{bc} F_{de}+3 \mathcal{F}_{bc}  \mathcal{F}_{de}\Big)\nonumber\\
&~~~~~~~~~~~~~~~~+\zeta  R^{m} _{~nbc}R^{n}_{~mde} \bigg)
-(D_a\Phi)^*(D^a\Phi)-V(\Phi)\bigg]\,,\label{eq:holomodel}
\end{align}
where $2\kappa^2$ is the five dimensional gravitational coupling constant, and $L$ is the AdS radius. We shall set  $2\kappa^2=L=1$ from now on. Two $U(1)$ gauge fields $V_a$ and $A_a$ are dual to the vector current and the axial current respectively, and their field strengths are
\be
\mathcal{F}_{ab}=\partial_a V_b-\partial_b V_a\,,~~~ F_{ab}=\partial_a  A_b-\partial_b A_a\,.
\ee
The Chern-Simons terms are introduced to give the correct anomalous structure for the dual field theory, including   $U(1)_A^3$ and $U(1)_AU(1)_V^2$ anomaly and mixed gauge-gravitational anomaly.
The scalar field $\Phi$ is axially charged and has the potential
\be D_a\Phi = (\partial_a - i q A_a)\Phi\,,~~V(\Phi)=m^2 |\Phi|^2 + \frac{\lambda}{2} |\Phi|^4\,. \ee
We will choose $m^2=-3, \lambda=1/2, q=4/5$ without loss of generality in this paper.\footnote{In section \ref{3.2} we shall switch the parameters $\lambda$ and $q$ while keep the scaling exponent fixed to study the universality of the anomalous transports after a proper renormalization in the critical region.\label{foot6}}

The ansatz for the background at finite temperature is
\be
ds^2=-udt^2+\frac{dr^2}{u}+ f(dx^2+ dy^2)+h dz^2\,,~~A=A_z dz\,,~~\Phi=\phi(r)\,,
\ee
where $u, f, h,A_z, \phi$ are functions of the radial coordinate $r$. The equations of motion for these fields can be found in \cite{Landsteiner:2015pdh}. From the conserved quantity $(\sqrt{h}(u'f-uf'))'=0$, one concludes $f=u$ at the zero temperature. This is also consistent with the intuition that there is an $SO(1,2)$ symmetry along $t, x, y$ directions for the zero temperature dual field theory. 
%At zero temperature, \red{$f=u$} \blue{one can find that $f=u$ due to the existence of a radially conserved quantity $(\sqrt{h}(u'f-uf'))'=0$} from the equation of motion\cite{Landsteiner:2015lsa}. 
Close to the UV boundary, we have $\lim_{r\to \infty}A_z=b$ and $\lim_{r\to\infty} r\phi=M$, which are the time-reversal symmetry breaking parameter and the axial symmetry breaking parameter separately. The details of the expansion for the background fields near the boundary can be found in appendix \ref{app:a}.

\begin{figure}[h!]
  \centering
  \includegraphics[width=0.600\textwidth]{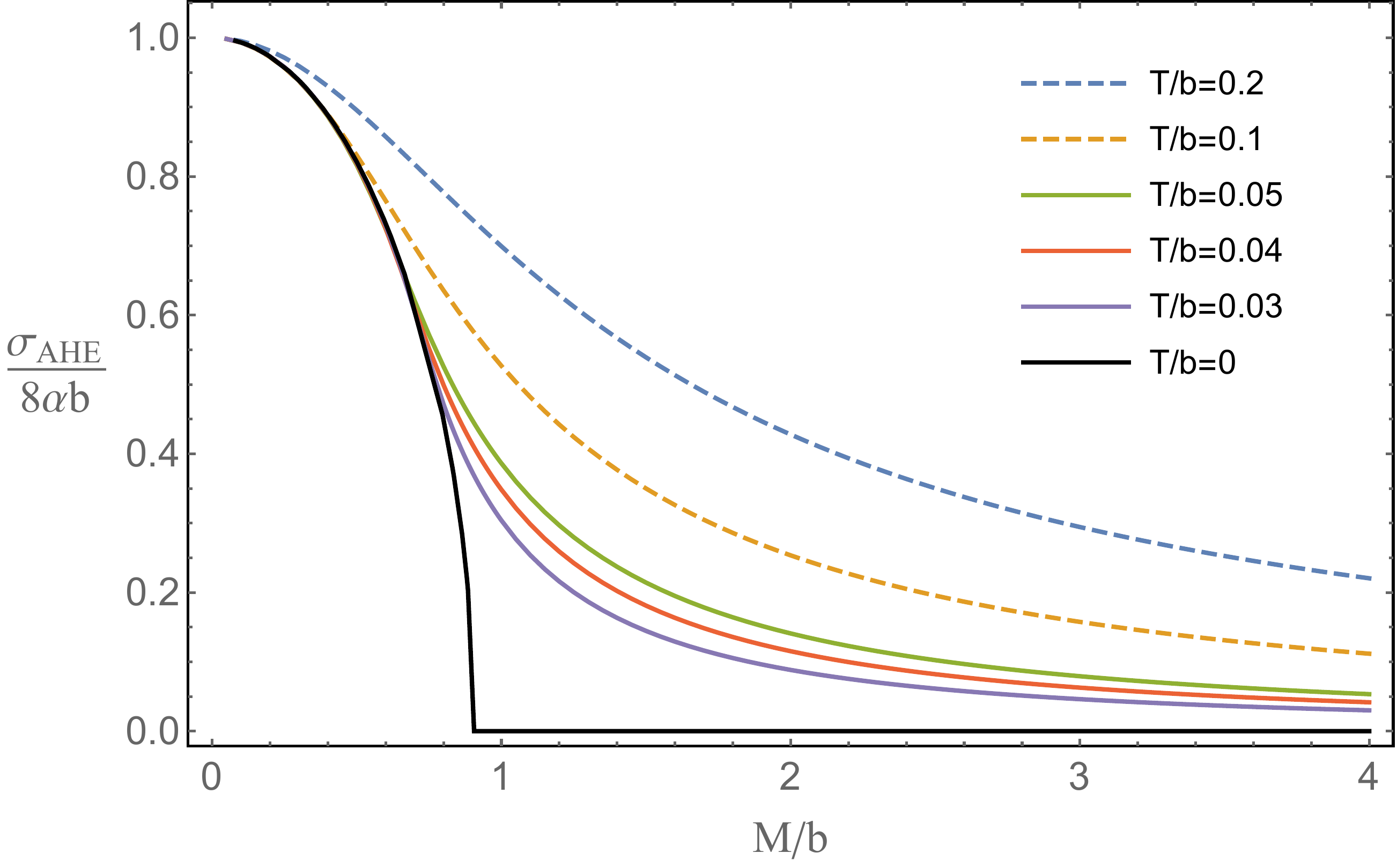}
  \caption{\small Anomalous Hall conductivity as a function of $M/b$ at zero and finite temperatures in the holographic Weyl semimetal for $m^2=-3, \lambda=1/2, q=4/5$. High ($T/b=0.2, 0.1$), low ($T/b=0.05, 0.04, 0.03$), and zero temperatures ($T/b=0$) are in dashed, solid and thick lines, respectively. }%For $T/b=0$, the discontinuity in $\sigma_\text{AHE}/b$ indicates that there is a quantum phase transition at the quantum critical point (QCP) $M_c/b=0.906$. At finite temperatures $T/b>0$, the QCP is enlarged to a quantum critical regime (QCR). As a result, the sharp line becomes smooth crossovers. In the WSM phase, $\sigma_\text{AHE}/b$ is independent of $T/b$ at low temperatures.}
  \label{fig:ahe}
\end{figure}

At zero temperature there exist three different kinds of near horizon solutions, which flow to three different regimes of $M/b$ at the UV boundary \cite{Landsteiner:2015pdh}.  
In the following, we briefly summarize these three kinds of IR solutions near the horizon, and highlight the emergent symmetry of the Lifshitz-type critical solution.
\\
\\
\textit{Weyl semimetal phase}: At the leading order the first type of IR solution 
 is an AdS$_5$ with non vanishing axial gauge field $A_z(r=0)$. Including the subleading correction, it 
takes the form
\be
u=r^2\,,~~h=r^2\,,~~A_z=a_1+\frac{\pi a_1^2\phi_1^2}{16r}e^{-\frac{2a_1q}{r}}\,,~~\phi =\sqrt{\pi}\phi_1\left(\frac{a_1q}{2r}\right)^{3/2}e^{-\frac{a_1q}{r}}\,,
\ee
where $a_1, \phi_1$ are the parameters for the IR geometry that are related to the parameters $M$ and $b$ for the UV geometry via RG flow. This type of IR geometry can flow to UV with $M/b<0.906$ for $ \lambda=1/2, q=4/5$. %This critical value is determined from the Lifshitz-type IR solution.
\\
\\\textit{Quantum critical point}: At the leading order the second type of IR solution is
\bea
ds^2&=&u_0r^2(-dt^2+dx^2+dy^2)+\frac{dr^2}{u_0r^2}+h_0r^{2\beta}dz^2\,,\nn\\
A_z&=&r^{\beta}\,,~~\phi=\phi_0\,.
\eea
With irrelevant perturbation, the IR geometry can flow to UV and it turns out in UV $(M/b)_c \simeq 0.906$, which corresponds to the critical point in the topological quantum phase transition \cite{Landsteiner:2015pdh}.
This critical solution has anisotropic Lifshitz-type symmetry under the scaling transformation $(t, x, y, r^{-1}) \rightarrow s(t, x, y, r^{-1}) $ and $z \rightarrow s^{\beta}z$.
There are four numerical values $\{u_0, h_0, \beta, \phi_0\}$ that are determined by the value of $\lambda, m$ and $q$. 
In particular, $\beta$ depends on two parameters $q, \lambda$ as 
\be
\beta=-\frac{2q^2}{-3-2q^2+\lambda \phi_0^2}\,,
\ee
where $\phi_0$ depends on $q, \lambda$ in a complicated way. Therefore, we can continuously tune $q$ and $\lambda$ to obtain different critical solutions which have the same emergent Lifshitz scaling exponent $\beta$. This makes it possible 
to study the relations between the anomalous transports and the emergent Lifshitz scaling exponent.
Furthermore, the null-energy condition constrains that $\beta\leq 1$ and regularity of solutions demands that $\beta> 0$, i.e., $\beta\in (0,1]$ \cite{Copetti:2016ewq}.
\\
\\\textit{Topological trivial phase}: The third type of IR solution at the leading order is an AdS$_5$ with non vanishing $\phi$. It takes the form
\be
u=(1+\frac{3}{8\lambda})r^2,~~h=r^2,~~A_z=a_1r^{\beta_1},~~\phi=\sqrt{\frac{3}{\lambda}}+\phi_1r^{\beta_2},
\ee
where $(\beta_1, \beta_2)=(\sqrt{1+\frac{48q^2}{3+8\lambda}}-1, 2\sqrt{\frac{3+20\lambda}{3+8\lambda}}-2)$.
This type of IR geometry can flow to the asymptotic AdS boundary with $M/b>0.906$.
\\
\\
The anomalous Hall conductivity in these three different regimes behaves as shown in the plot of  Fig. \ref{fig:ahe}. For $T/b=0$, the discontinuity of $\sigma_\text{AHE}/b$ indicates that there is a quantum phase transition at the quantum critical point (QCP) $M_c/b\blue ~\simeq ~0.906$. At finite temperatures $T/b>0$, the QCP is enlarged to a quantum critical region (QCR). As a result, the sharp line becomes smooth crossovers. In the Weyl semimetal phase, $\sigma_\text{AHE}/b$ is independent of $T/b$ at low temperatures.

%%%%%%%%%%%%%%%%%%%%%%%%%%%%
%%%%%%%%%%%%%%%%%%%%%%%%%%%%

%%%%%%%%%%%%%%%%%%%%%%%%%%
\section{Chiral vortical conductivity and the universality}
\label{sec3}
%%%%%%%%%%%%%%%%%%%%%%%

In this section, we will study the chiral vortical conductivity at finite temperature in the holographic model (\ref{eq:holomodel}). We will first focus on the crossover of chiral vortical conductivity across the topological phase transition at different temperatures. From field theory point of view this is %analogous
to investigate how the fermion mass influences the chiral vortical effect in a time reversal symmetry breaking system. Note that, we will concentrate on the contribution induced from the mixed gauge-gravitational anomaly. %Therefore, the (axial) chemical potential will be turned off in order to clearly demonstrate the temperature dependence of the chiral vortical conductivity. %as another interest in \ref{subsec3.1}.
In Sec. \ref{3.2}, we will study the physics in the critical region, including the behaviors of  the chiral vortical conductivity and odd (Hall) viscosities. %There is a subtle point that
Since
different critical ground states could have same emergent scaling exponent $\beta$ of the Lifshitz symmetry, as explained at the footnote \ref{foot6}, we will verify the possible existence of universal physics in the critical region.
%The ansatz for the anisotropic background at finite temperature is
%\be
%ds^2=-udt^2+\frac{dr^2}{u}+ f(dx^2+ dy^2)+h dz^2\,,~~A=A_z dz\,,~~\phi=\phi(r)\,,
%\ee
%where fields $u, f, h, A_z, \phi$ are functions of the radial coordinate $r$. For zero temperature, $u=f$.
%\red{At finite temperature, the black hole horizon is located at $r_0$ where $u(r_0)=0$. One can always set $r_0=1$ using the translational symmetry of the radial coordinate.  }

%%%%%%%%%%%%%%%%%%%%%%%%%%%%%%
\subsection{Chiral vortical conductivity}
\label{subsec3.1}
%%%%%%%%%%%%%%%%%%%%%%%%%%%%%%

The chiral vortical conductivity for the dual field theory of the holographic model can be computed from the Kubo formula
\be\label{eq:kubofor}
\sigma_{V}=\mathop{\lim}\limits_{k_l\rightarrow0}\frac{i}{2k_l}\sum_{m,n}\epsilon_{lmn}\langle J_5^nT^{tm}\rangle\Big{|}_{\omega=0},
\ee
where $m,n,l\in\{x,y,z\}$, $J_5$ is the axial current and $T^{\mu\nu}$ is the energy momentum tensor. The holographic Weyl semimetal is an anisotropic system and we have two independent chiral vortical conductivities. For simplicity, we shall focus on the component with fluctuation along the $z$-direction,\footnote{To consider the other component of the chiral vortical conductivity, we need to turn on all the fluctuations around the background to do the calculation. This is extremely complicated and beyond the scope of this work.} i.e. we choose  $k_z\to 0$ limit in (\ref{eq:kubofor}). %and we have chosen the coordinates satisfying $\vec{b}=b\hat{e}_z$, ...we choose...

We switch on the following perturbations
\be \delta g_{ti}=h_{ti}e^{ik_zz}\,, ~~\delta g_{iz}=h_{iz}e^{ik_zz}\,,~~\delta A_i=a_ie^{ik_zz}\ee
for $i\in \{x, y\}$. With six perturbation fields in consideration, one obtains rather intractable equations of motion for $\delta A_i$ as the corresponding response to the source $\delta g_{tj}$.
The details of equations can be found in appendix \ref{app:cve}.
For all the fields we expand as e.g. $a_i=a_i^{(0)}+k_z a_i^{(1)}+\dots$, and we solve the equations order by order in $k_z$. At zeroth order, the solutions are $h_{ti}^{(0)}=u$ and $a_i^{(0)}=0$. At first order, $h_{ti}^{(1)}=0$ and
%utilising the constraint equation () and the sourceless boundary condition,
the equation of motion for $a_i^{(1)}$ is %to the lowest order in $k_z$ can be written as
\bea\label{eq:1sta}
a_{i}^{(1)\prime\prime}+\left(\frac{u^{\prime}}{u}+\frac{h^{\prime}}{2h}\right)a_{i}^{(1)\prime}-\left(\frac{2q^{2}\phi^{2}}{u}-\frac{A_z'^{2}}{h}\right)a_{i}^{(1)}
-\frac{2i\zeta }{\sqrt{h}}\epsilon_{ij}\left(\frac{u'}{u}-\frac{f'}{f}\right)L_1&=&0\,,
\eea
with
\be
L_1=\left[-\left(\frac{2h'}{h}+\frac{f'}{f}\right)h_{tj}^{(0)\prime}+\left(\left(\frac{f^{\prime\prime}}{f}-\frac{h^{\prime\prime}}{h}\right)+
\frac{3h^{\prime}}{2h}\left(\frac{f^{\prime}}{f}+\frac{h^{\prime}}{h}\right)+\frac{A_z'^2}{h}\right)h_{tj}^{(0)}\right]\,,
\ee
where $\epsilon_{xy}=-\epsilon_{yx}=1$. Note that this equation of motion decouples from $h_{iz}$. We make some observations here before discussing the numerical results:
\begin{itemize}
\item When $T=0$, $u=f$, the chiral vortical conductivity vanishes. In other words, at zero temperature, the rotation  cannot induce any axial current.
\item At finite temperature $\sigma_V\propto \zeta$. This indicates that the chiral vortical effect in this system is completely induced from the mixed gauge-gravitational anomaly.
\item When $\phi=0$, we have $A_z=b$, $f=h=r^2$, and $u=r^2\big(1-\frac{r_0^4}{r^4}\big).$ This is exactly the Schwarzschild background with a constant $A_z$.  The equation of motion (\ref{eq:1sta}) can be solved analytically. From the equation (\ref{eq:1sta}), one obtains
\be
a_i^{(1)}=\int_{r_0}^{r}d\tilde{r}\left[\frac{32i\zeta r_0^8}{u\sqrt{h}\tilde{r}^6}-\frac{32i\zeta r_0^2}{u\sqrt{h}}\right].
\ee
Therefore, $a_i^{(1)}=\frac{16i\zeta r_0^2}{r^2}$ near the boundary with $r_0=\pi T$ and one obtains $\sigma_V=32 \pi^2 \zeta T^2$. This is exactly the same as results in \cite{Landsteiner:2011cp, Landsteiner:2011iq} in the zero density limit.
\item For the generic case with $\phi\neq0$, there is not an analytical bulk solution. The numerical tool is necessary to obtain the chiral vortical conductivity.
\end{itemize}

The numerical approach to solving (\ref{eq:1sta}) is summarized as follows. First, we numerically obtain the background solution \cite{Landsteiner:2016stv}. Starting from the near horizon geometry, one can integrate to the boundary to obtain the whole geometry in the bulk. Due to the scaling symmetries, there are two free shooting parameters, which correspond to $M/b$ and $T/b$ in the dual field theory. Then with the background solution obtained for given $M/b$ and $T/b$, the equation (\ref{eq:1sta}) can be solved with the appropriate regular boundary condition at the horizon and the sourceless boundary condition at the conformal boundary. The details to obtain the Green's function can be found in appendix \ref{app:cve}.

For convenience, we define the dimensionless mass, the dimensionless temperature and the dimensionless (normalized) conductivities as
\be
\hat{M}\equiv \frac{M}{b}\,,~~~\hat{T}\equiv \frac{T}{b}\,,~~~
\tilde{\sigma}_\text{AHE}\equiv\frac{\sigma_\text{AHE}}{8\alpha b}\,,~~~~\tilde{\sigma}_{V}\equiv\frac{\sigma_{V}}{32\pi^2 \zeta b^2}\,.
\ee

\begin{figure}[h!]
  \centering
  \includegraphics[width=0.600\textwidth]{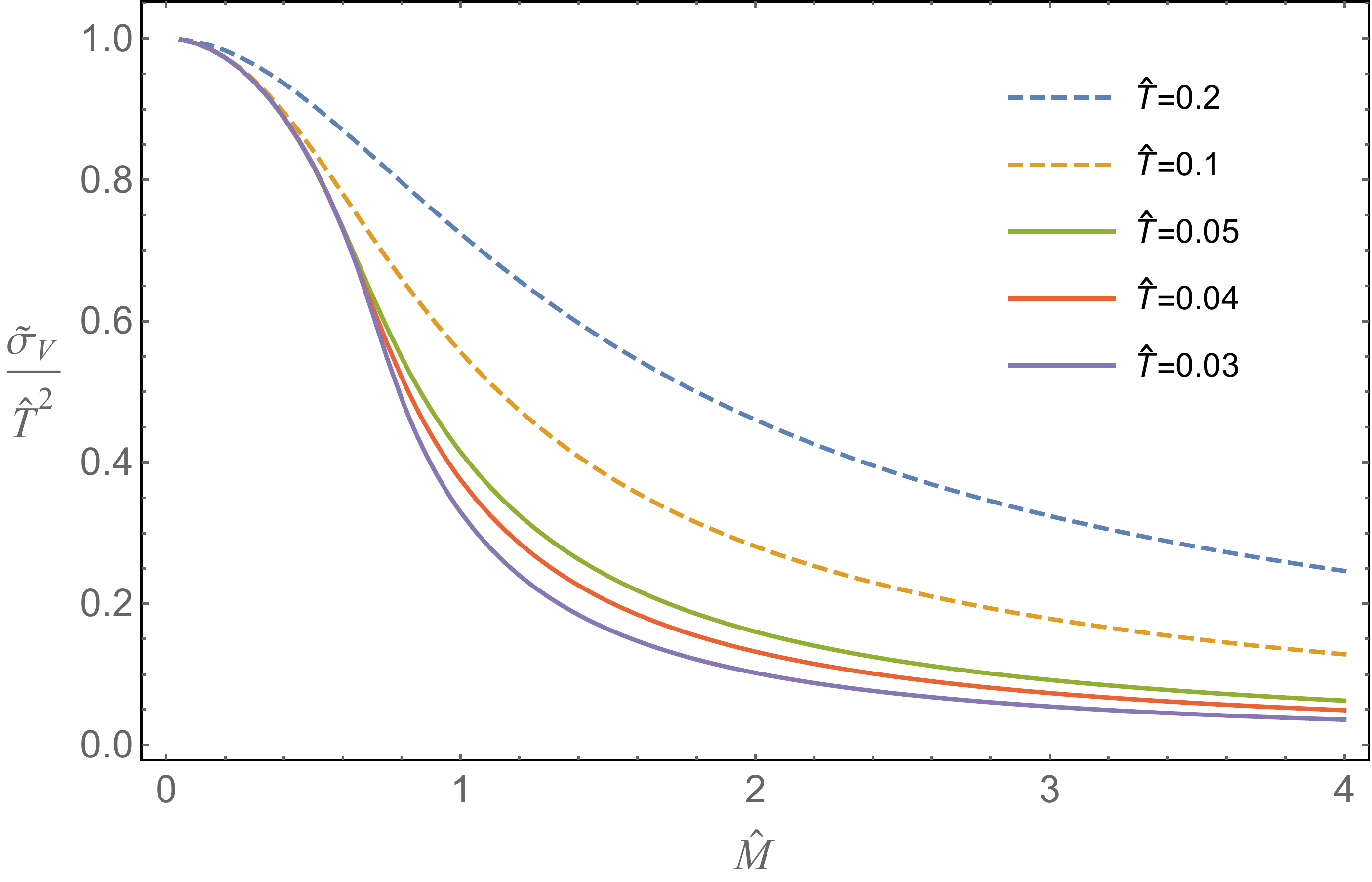}
  \caption{\small The dependence of the chiral vortical conductivity $\tilde{\sigma}_V$ over $\hat{T}^2$ on $\hat{M}$ at different temperatures. $\tilde{\sigma}_V/\hat{T}^2$ exhibit similar behavior with respect to $\tilde{\sigma}_\text{AHE}$ as a function of dimensionless $\hat{M}$.}
 % At low temperatures($\hat{T}\leq0.05$), $\tilde{\sigma}_V/\hat{T}^2$ at various temperatures coincide before reaching the quantum critical region(QCR), which indicates a $\hat{T}^2$ power law of this transport in the WSM phase. At large $\hat{M}\gg\hat{M}_c$ in the topological trivial phase, $\tilde{\sigma}_V$ tend to vanish as a result of the chiral fermion degrees of freedom being gapped out. A crucial difference between $\tilde{\sigma}_V$ and $\tilde{\sigma}_{AHE}$ is at $\hat{T}=0$, $\tilde{\sigma}_{AHE}$ shows a sharp but continuous phase transition at the QCP, while $\tilde{\sigma}_V$ vanishes in all these three phases. Note that, although at large $\hat{M}\sim10^2$ both $\tilde{\sigma}_V$ and $\tilde{\sigma}_{AHE}$ approach $0$, the ratio of these two conductivities shows a nontrivial behavior, as shown in Fig. \ref{fig:ncveM}.}
  \label{fig:cveM}
\end{figure}

In Fig. \ref{fig:cveM}, we numerically demonstrate the reduced chiral vortical conductivity over temperature squared $\tilde{\sigma}_V/{\hat T}^2$ as a function of $\hat{M}$ at various temperatures. When $\hat{M}\rightarrow0$, one obtains $\sigma_V=32\pi^2\zeta T^2$ which is the case that we do not explicitly break the chiral symmetry \cite{Landsteiner:2011cp, Landsteiner:2011iq}.   When $\hat{M}$ increases, the chiral vortical conductivity decreases. Within the Weyl semimetal phase, as the temperature is sufficiently low ($\hat{T}\leq0.05$), $\tilde{\sigma}_V/\hat{T}^2$ shows a perfect coincidence, similar to the anomalous Hall conductivity in Fig \ref{fig:ahe}. This  means that $\tilde{\sigma}_V\propto\hat{T}^2$ in the Weyl semimetal phase. In the quantum critical region and the topologically trivial phase, the chiral vortical conductivity decreases monotonically and approaches zero as a result of the degrees of freedom of chiral fermions being gradually gapped out. It is interesting to note that Fig. \ref{fig:cveM} shows the mass dependence of $\tilde{\sigma}_V/\hat{T}^2$ of the strongly coupled massive fermion system, which behaves qualitatively the same as
the investigation of mass effect of chiral vortical effect in the weakly coupled massive fermion system \cite{Lin:2018aon, Flachi:2017vlp}. We should emphasize that in our studies the time reversal symmetry is broken while this symmetry is preserved in \cite{Lin:2018aon, Flachi:2017vlp}.

\vspace{.3cm}
%%%%%%%%%%%%%%%%%%%%%%%%%%%%
%%%%%%%%%%%%%%%%%%%%%%%%%%%%
%At low temperatures($\hat{T}\leq0.05$), $\tilde{\sigma}_V/\hat{T}^2$ at various temperatures coincide before reaching the quantum critical region(QCR), which indicates a $\hat{T}^2$ power law of this transport in the WSM phase. At large $\hat{M}\gg\hat{M}_c$ in the topological trivial phase, $\tilde{\sigma}_V$ tend to vanish as a result of the chiral fermion degrees of freedom being gapped out.

%A crucial difference between $\tilde{\sigma}_V$ and $\tilde{\sigma}_{AHE}$ is at $\hat{T}=0$, $\tilde{\sigma}_{AHE}$ shows a sharp but continuous phase transition at the QCP, while $\tilde{\sigma}_V$ vanishes in all these three phases. Note that, although at large $\hat{M}\sim10^2$ both $\tilde{\sigma}_V$ and $\tilde{\sigma}_{AHE}$ approach $0$, the ratio of these two conductivities shows a nontrivial behavior, as shown in Fig. \ref{fig:ncveM}.

\begin{figure}[h!]
  \centering
\includegraphics[width=0.35\textwidth]{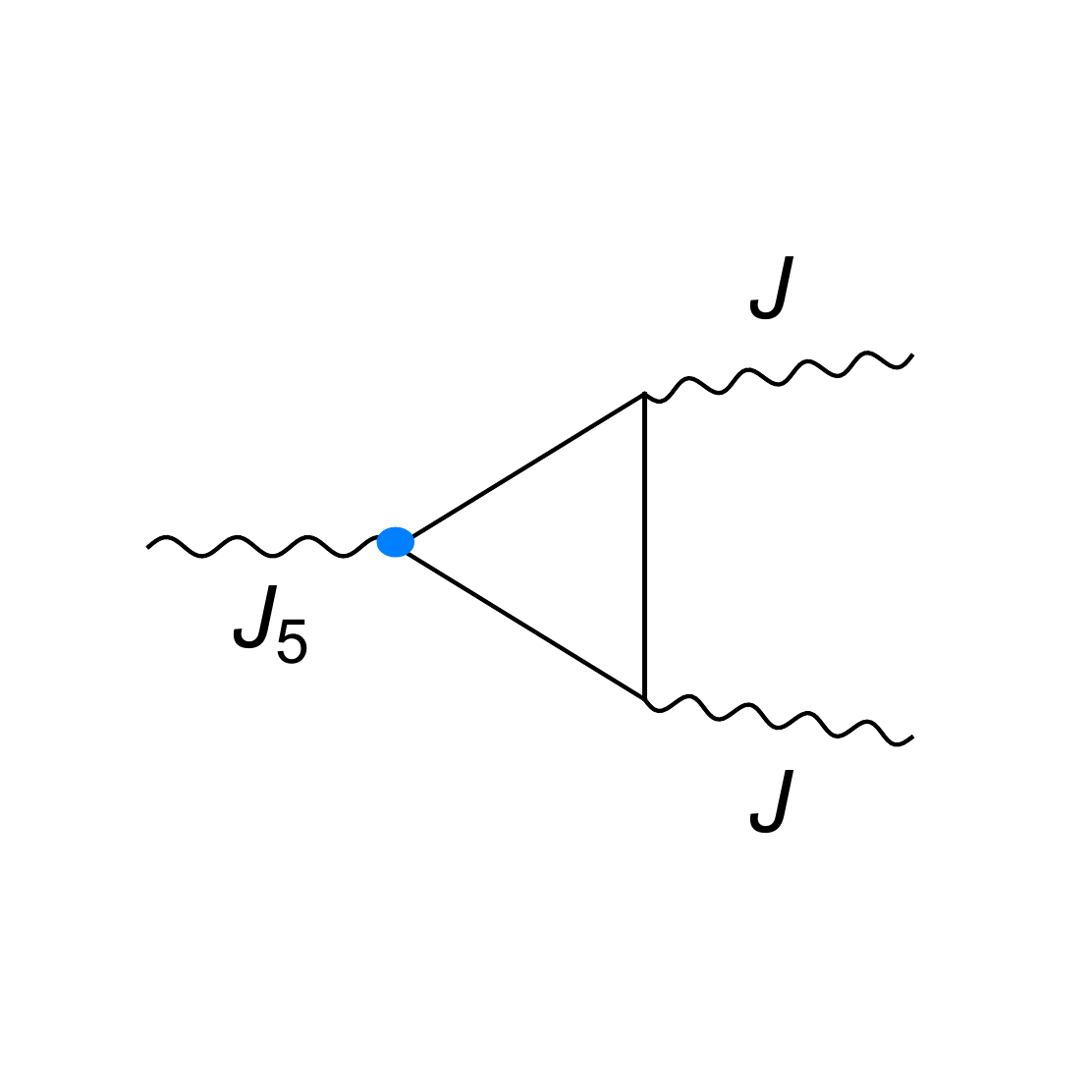}
\includegraphics[width=0.32\textwidth]{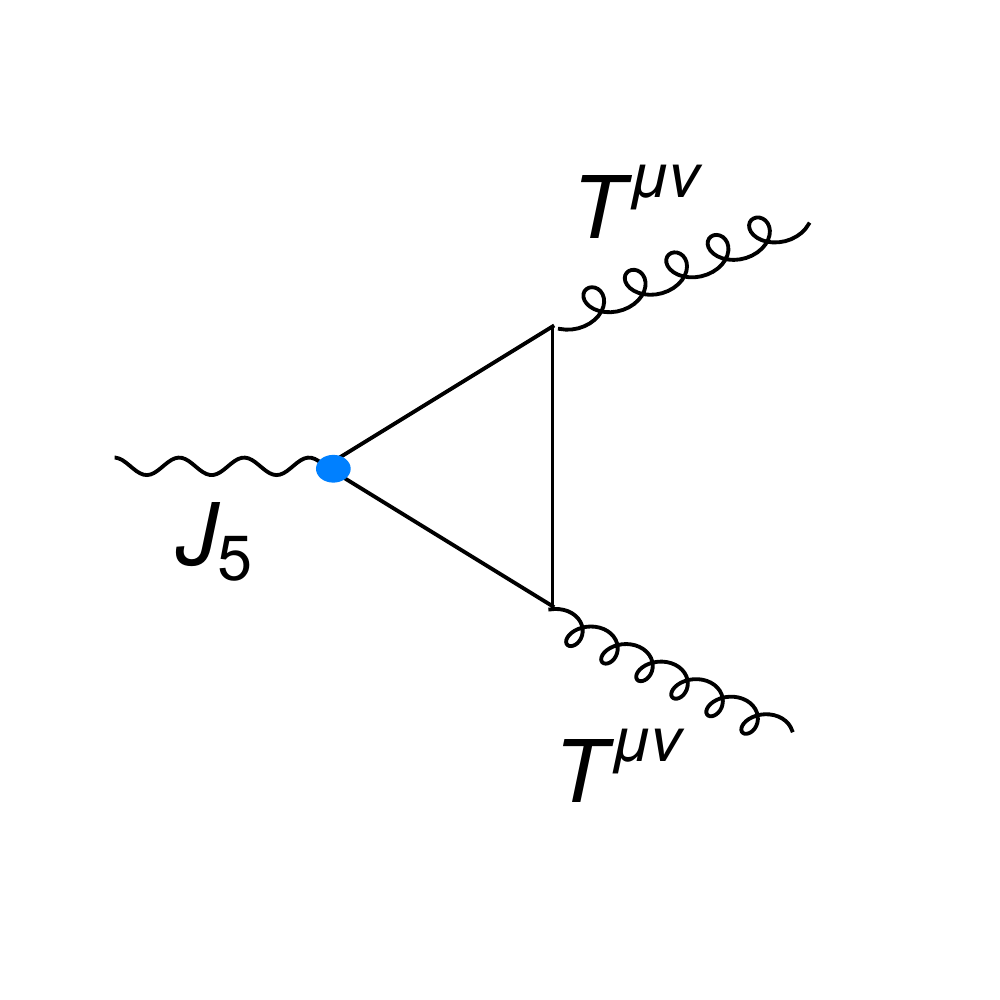}
\vspace{-1.2cm}
  \caption{\small Feynman diagrams of the leading-order corrections to the two-point correlators $\langle JJ\rangle$ and $\langle J_5T\rangle$. These are the most significant contributions when computing the anomalous Hall (the left plot) and the chiral vortical (the right plot) conductivities. The blue dot in each diagram represents the renormalized factor that is proportional to $\sqrt{Z_A}$. Both the two diagrams scale as $\sqrt{Z_A}$.}
  \label{fig:ren}
\end{figure}

Moreover, from Figs. \ref{fig:ahe} and \ref{fig:cveM} one may notice that $\tilde{\sigma}_V/\hat{T}^2$ shows very similar behavior compared to $\tilde{\sigma}_\text{AHE}$.\footnote{Note that there is one obvious difference that at zero temperature
$\tilde{\sigma}_V$ vanishes in all the three phases while $\tilde{\sigma}_\text{AHE}$ does not vanish in the Weyl semimetal phase. } Although both of these two conductivities approach zero at large $\hat{M}\gg\hat{M}_c$, the ratio of these two conductivities shows nontrivial physics. The reason to investigate the normalised ratio of $\tilde{\sigma}_V/\hat{T}^2$ to $\tilde{\sigma}_\text{AHE}$ is motivated by the suggestion made in \cite{Copetti:2016ewq} that when we compute the two point correlation functions, proper renormalization effect of the axial current should be taken into account. In other words, when we consider the physics at low energy scale near the Fermi surface, the DC conductivities will be related to the IR coupling of
the external fields, which is in general different from the UV strength. Following \cite{Copetti:2016ewq} we introduce a renormalized factor $\sqrt{Z_A}$ satisfying
\begin{equation}\label{renormalized constant}
\sqrt{Z_A}b=b^{\text{IR}}\,,
\end{equation}
where
\begin{equation}\label{buv&bir}
b=\mathop{\lim}\limits_{r\rightarrow\infty}A_z\,, ~~~~b^{\text{IR}}=A_z(r_0)\,.
\end{equation}
This coupling renormalization has impacts both on the anomalous Hall conductivity $\tilde{\sigma}_\text{AHE}$ and the chiral vortical conductivity $\tilde{\sigma}_V$, as shown in the leading order contribution of Feynman diagrams for the anomalous contributions to the two-point correlators $\langle JJ\rangle$ and $\langle J_5T\rangle$ in Fig. \ref{fig:ren}.
\vspace{0cm}

\begin{figure}[h!]
  \centering
  \includegraphics[width=0.600\textwidth]{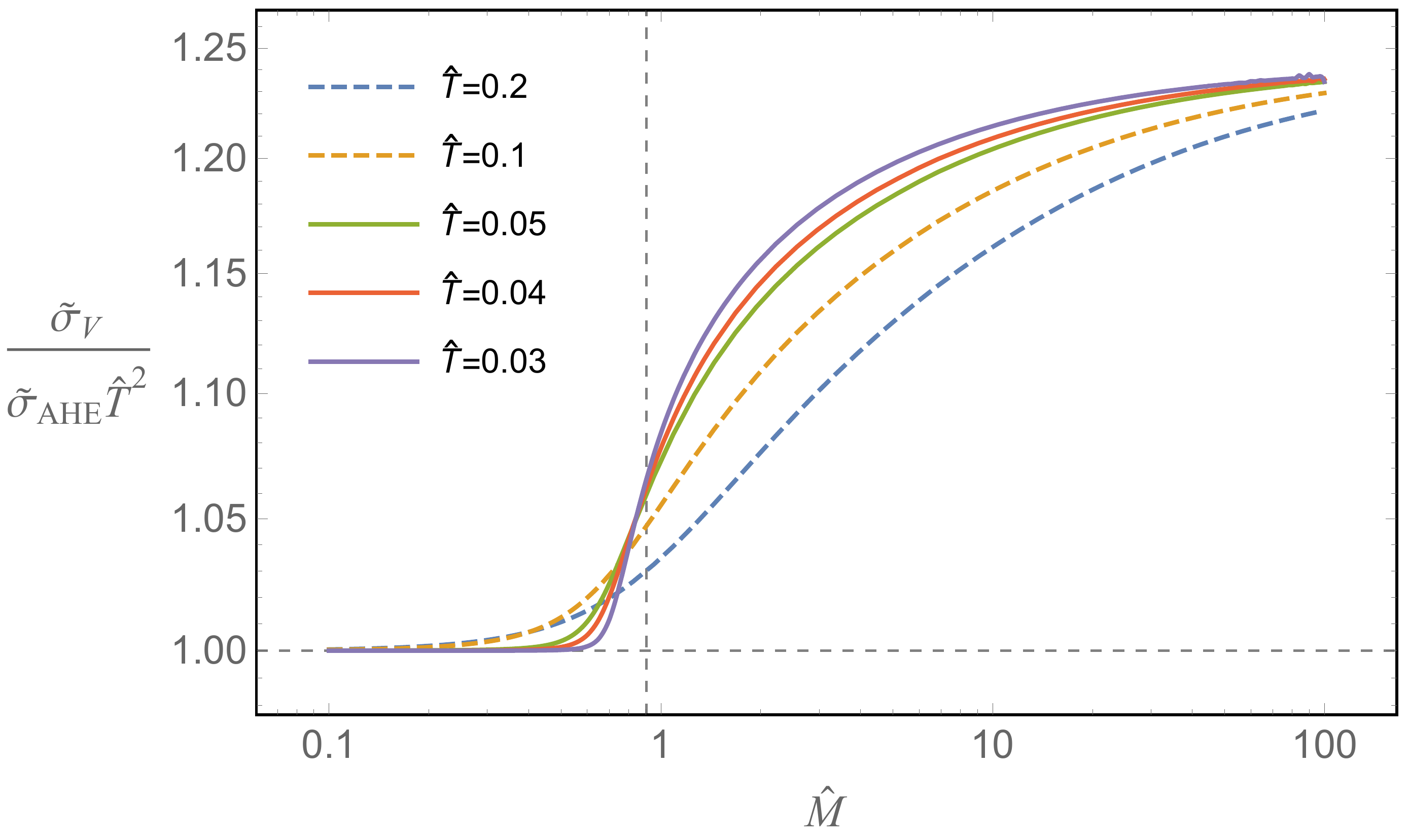}
  \caption{\small The normalized chiral vortical conductivity $\tilde{\sigma}_V$ over normalized anomalous Hall conductivity $\tilde{\sigma}_\text{AHE}$ and $\hat{T}^2$ as a function of dimensionless $\hat{M}$ at different temperatures. The perpendicular grid-line at $\hat{M}_c=0.906$ indicates the QCP. High temperatures ($\hat{T}=0.2, 0.1$) and low temperatures ($\hat{T}=0.05,0.04,0.03$) are in dashed and solid lines, respectively. At low temperature the ratio departs from a constant value in the Weyl semimetal phase, increases in the vicinity of the QCP and reaches a temperature-independent value at large $\hat{M}$.}
  %As one lowers the temperature, the ratios are almost a constant in the WSM phase as a result of infrared screening; the ratios increase in the vicinity of the QCP and reach a temperature-independent value at large $\hat{M}$. Note that $\tilde{\sigma}_V/\tilde{\sigma}_\text{AHE}\propto \hat{T}^2$ is also satisfied in the quantum critical region and further, there exist universal behaviors for the mixed anomaly induced transports in this phase, as explained in detail in Sec. \ref{3.2}.}
\label{fig:ncveM}
\end{figure}

Therefore, to uncover the possible universal physics at low energy, we renormalize the chiral vortical conductivity through $\tilde{\sigma}_\text{AHE}$. The behavior of $\tilde{\sigma}_V/(\tilde{\sigma}_\text{AHE}\hat{T}^2)$ as a function of $\hat{M}$ is shown in Fig. \ref{fig:ncveM}, from which several universalities  regarding to chiral vortical effect could be revealed. Firstly, in the Weyl semimetal phase at low temperatures, the ratio is almost a constant
due to the cancellation of infrared screening in both these two transports. Although with different choices of $\hat{M}$ the underlying bulk geometries have different profiles, this ratio is in coincidence with the case $\hat{M}=0$. More interestingly, this property is the same in the topological nontrivial phase for all possible choices of ($q$, $\lambda$) that allow a quantum phase transition in this holographic model.
Secondly, in the quantum critical region $\tilde{\sigma}_V/\tilde{\sigma}_\text{AHE}\propto \hat{T}^2$. More universal behaviors for the mixed anomaly induced transports in this phase will be studied in detail in Sec. \ref{3.2}.
Thirdly, at large enough $\hat{M}\gg\hat{M}_c$, the ratio approaches  another value, which is independent of $\hat{T}$
 and is greater than one while this value depends on the choice of ($q$, $\lambda$).

% Another interesting fact in the topological trivial phase is that, (diagram of $T^{\beta1}$)....
\vspace{.3cm}

\subsection{Universality of Mixed Anomaly Induced Transports}
\label{3.2}
%\red{In the previous subsection, we found after proper normalization of chiral vortical conductivity universal behavior of the renomalized chiral vortical conductivity can be studied.}
In this subsection, we will focus on the universal physics of the mixed gauge-gravitational anomaly induced transports in the quantum critical region.

\begin{figure}[h!]
  \centering
  \includegraphics[width=0.550\textwidth]{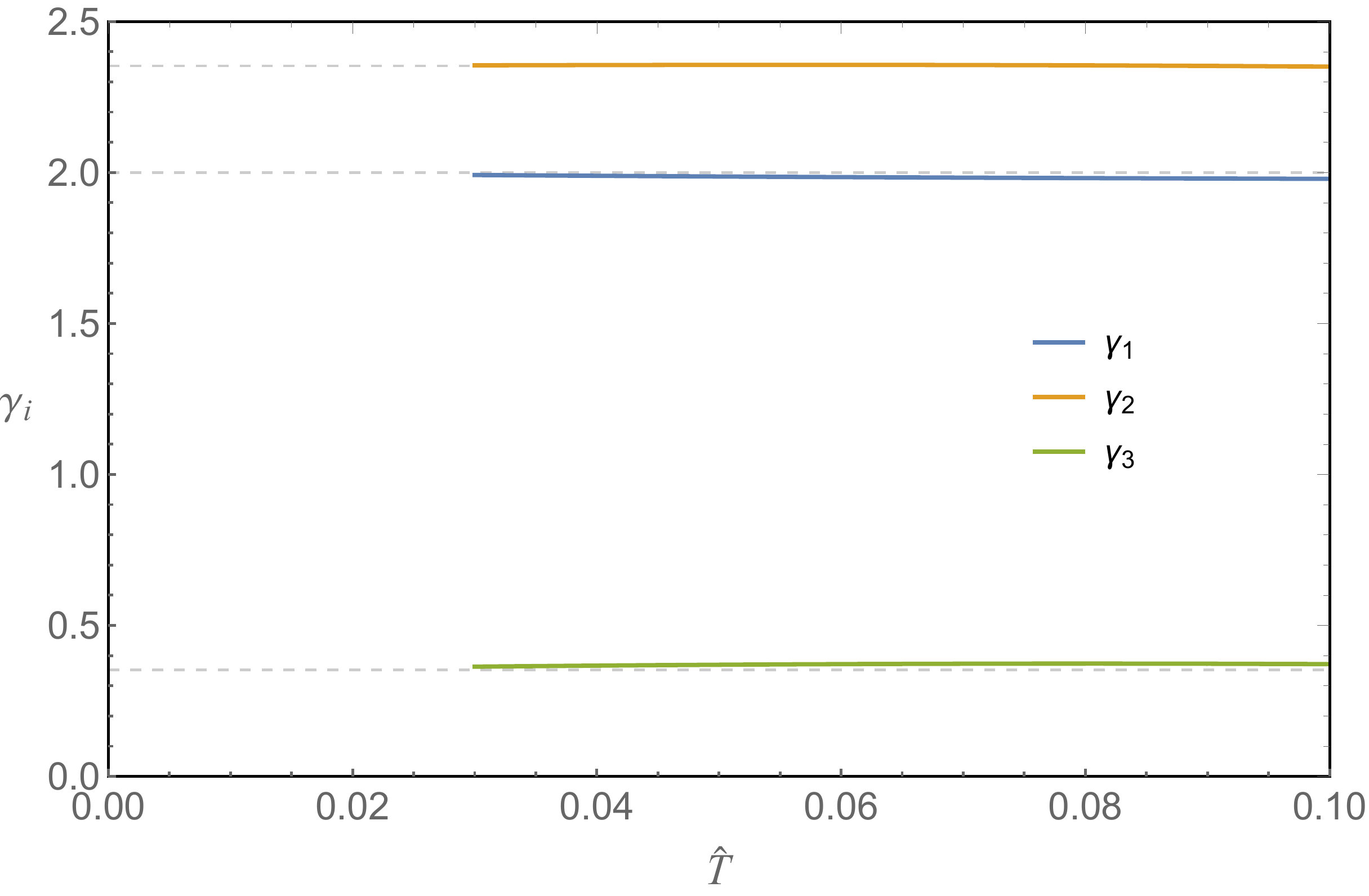}
  \caption{\small The temperature scaling exponents $\gamma_i$ with $\{\gamma_1, \gamma_2,\gamma_3\}$ for $\{\tilde{\sigma}_V/\tilde{\sigma}_\text{AHE},\tilde{\sigma}_V,\tilde{\sigma}_\text{AHE}\}$ respectively at the critical value $\hat{M}_c=0.906$ in the low-temperature regime. The three dashed lines from top to bottom denote $\gamma=2+\beta, 2$ and $\beta$ with $\beta=0.353$.}
  \label{fig:power}
\end{figure}

\begin{figure}[h!]
  \centering
  \includegraphics[width=0.600\textwidth]{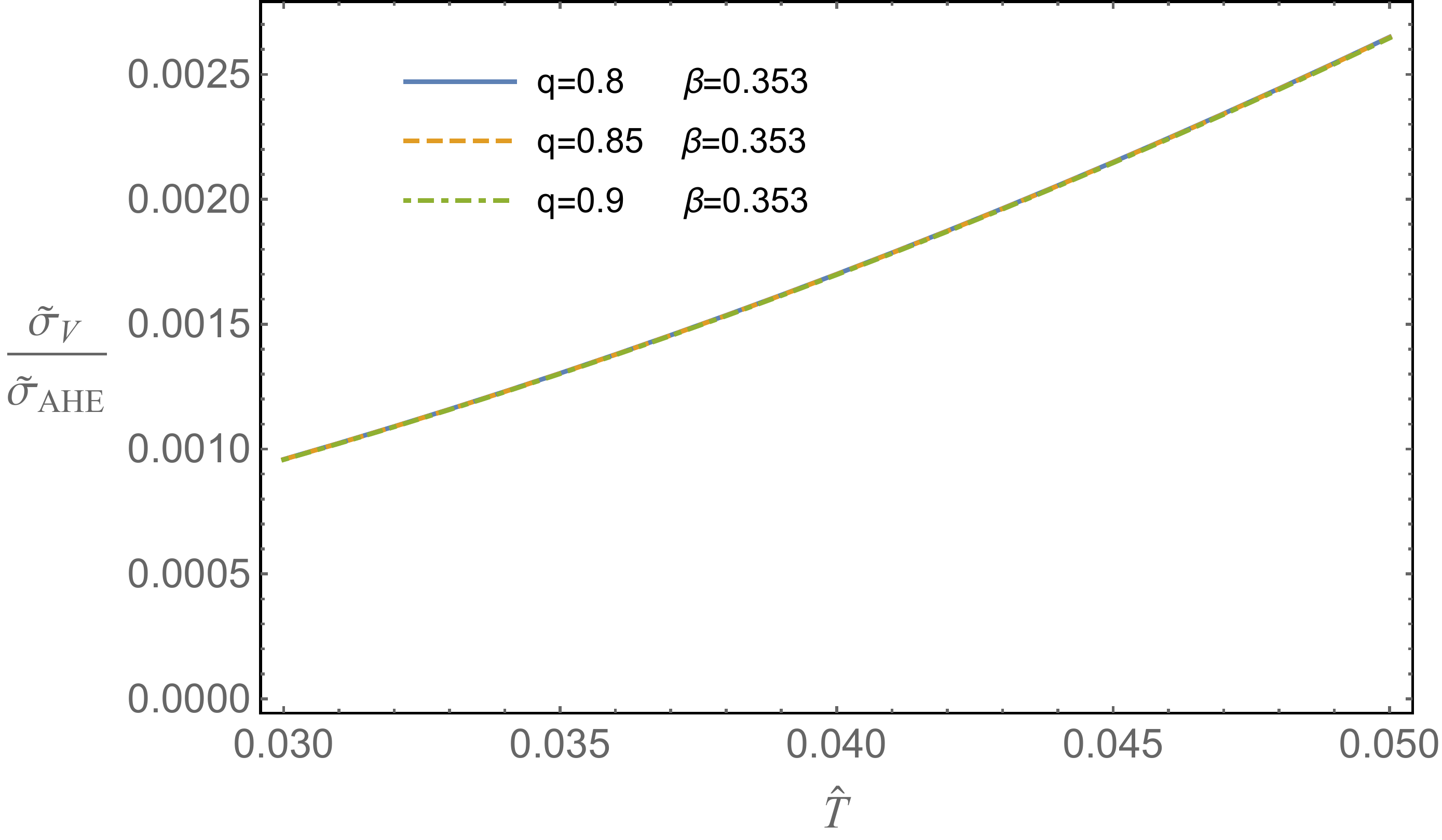}
  \caption{\small Universality of $\tilde{\sigma}_V/\tilde{\sigma}_\text{AHE}$ at low temperatures in the quantum critical region   with the IR scaling exponent $\beta=0.353$. With $\beta$ fixed, the choices of ($q$, $\lambda$) are (0.8, 0.5) (blue), (0.85, 0.308) (orange-dashed), (0.9, 0.027) (green-dotdashed) with $\hat{M}_c=0.906,~0.808, ~0.730$, respectively. However, $\tilde{\sigma}_V/\tilde{\sigma}_\text{AHE}$ is independent of the choice of ($q$, $\lambda$) and is only a function of $\beta$ and $\hat{T}$.}
  \label{fig:ncveT}
\end{figure}

As shown in Fig. \ref{fig:ncveM}, in the vicinity of the QCP located at $\hat{M}_c\simeq0.906$, $\tilde{\sigma}_V/(\tilde{\sigma}_\text{AHE}\hat{T}^2)$ tends to exhibit overlaps for a variety of low temperatures. One can clearly find the temperature-dependence of the renormalized chiral vortical conductivity from Fig. \ref{fig:ncveT}. As is adequately discussed in \cite{Landsteiner:2016stv}, at low temperature,  the scaling behavior of $\tilde{\sigma}_\text{AHE}$ in the quantum critical region is determined from the anisotropic scaling exponent $\beta$, i.e., $\tilde{\sigma}_\text{AHE}\propto \hat{T}^\beta$. However, for $\tilde{\sigma}_V$, we cannot analytically express it in terms of the horizon data due to the nonexistence of a radially  conserved quantity.  Nevertheless, a careful analysis of numerical data shows that, there exists an exact power law for the normalized chiral vortical conductivity when $\hat{T}\rightarrow0$, i.e., $\tilde{\sigma}_V\propto \hat{T}^{\gamma_2}$. In Fig. \ref{fig:power}, we show the temperature scaling exponents ${\gamma_1, \gamma_2,\gamma_3}$ of $\tilde{\sigma}_V/\tilde{\sigma}_{\text{AHE}}, \tilde{\sigma}_V,$ and $\tilde{\sigma}_{\text{AHE}}$ and find that these three exponents nicely match with  $(\gamma_1, \gamma_2,\gamma_3)=(2, 2+\beta, \beta)$.

%%%%%%%%%%%%%%%%%%%%%%%%
\vspace{0.3cm}
%%%%%%%%%%%%%%%%%%%%%%%%

%%%%%%%%%%%%%%%%%%%%%%%%
\vspace{.3cm}
%%%%%%%%%%%%%%%%%%%%%%%%

We also find that the renormalized chiral vortical conductivity exhibit universal behaviors in the quantum critical region as shown in Fig. \ref{fig:ncveT}. When we fix the IR scaling exponent $\beta=0.353$ without loss of generality, and change the two parameters $q$ and $\lambda$ in this holographic model, the locations of the QCPs $\hat{M}_c$ are different. However, $\tilde{\sigma}_V/\tilde{\sigma}_\text{AHE}$  in the cases with different $(q,\lambda)$ while same $\beta$ show exactly the same dependence on $\hat{T}$ at low temperature, while neither $\tilde{\sigma}_V$ nor $\tilde{\sigma}_\text{AHE}$ alone %do\red{es} not
behave in the same way. Therefore, numerical evidences strongly suggest that
\be\label{eq:cbeta}
\frac{\tilde{\sigma}_V}{\tilde{\sigma}_\text{AHE}}=c(\beta)\hat{T}^2
\ee
in the quantum critical region, where $c(\beta)$ is introduced as a coefficient function of the scaling exponent $\beta$.  %We numerically compute $c(\beta)$ in the low temperature regime.

\begin{figure}[h!]
  \centering
  \includegraphics[width=0.600\textwidth]{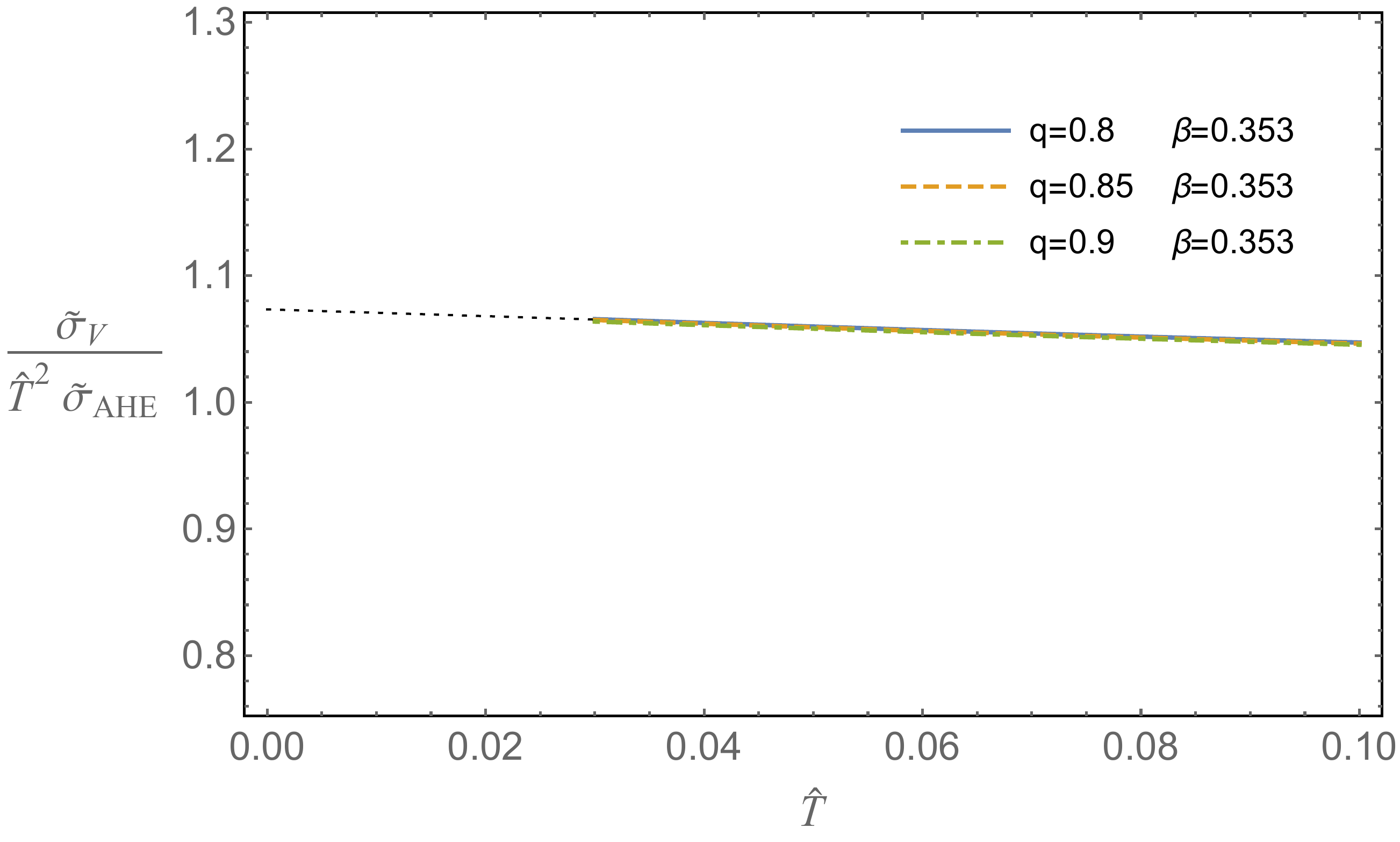}
  \caption{The coefficient function $c(\beta)$ at low temperatures for $\beta=0.353$. One finds that at low temperatures $c(\beta=0.353)$ is almost a constant. The black dashed line is the interpolation of $c(\beta)$ to $\hat{T}=0$ from which we can get the numerical error between $c|_{\hat{T}=0.03}$ and $c|_{\hat{T}\rightarrow0}$.}
  \label{fig:c(0.353)}
\end{figure}

With fixed $\beta=0.353$, one can study the behavior of $c(\beta)$ at low temperatures as shown in Fig. \ref{fig:c(0.353)}. Due to numerical difficulties, the temperature
$\hat{T}$  can not be arbitrarily small.
Fig. \ref{fig:c(0.353)} shows that this coefficient function $c(\beta)$ changes very slowly with a negative slope at about
$\Delta c/\Delta \hat{T} \approx -0.27$.
From a linear interpolation as shown by the black dashed line in Fig. \ref{fig:c(0.353)} one can read $c|_{\hat{T}\rightarrow0}$ from which one gets the numerical difference between $c|_{\hat{T}=0.03}$ and $c|_{\hat{T}\rightarrow0}$
\footnote{This coefficient function is not well-defined at zero temperature where chiral vortical conductivity vanishes. Also, $c|_{\hat{T}\rightarrow 0}$ is obtained from fitting the existing data as shown by the black dashed line in Fig. \ref{fig:c(0.353)}.}
\be\label{eq:err}
\frac{c|_{\hat{T}=0.03}-c|_{\hat{T}\rightarrow0}}{c|_{\hat{T}=0.03}}\simeq -0.8\%
\ee
for $\beta=0.353$. We have checked that the numerical error is always small for $\beta\in(0,1)$. This indicates that it is reasonable to obtain this coefficient function $c(\beta)$ at $\hat{T}=0.03$, given that the solutions are inaccessible at too low temperatures.

%%%%%%%%%%%%%%%%%%%%%%%%
\vspace{.3cm}
%%%%%%%%%%%%%%%%%%%%%%%%

\begin{figure}[h!]
  \centering
  \includegraphics[width=0.600\textwidth]{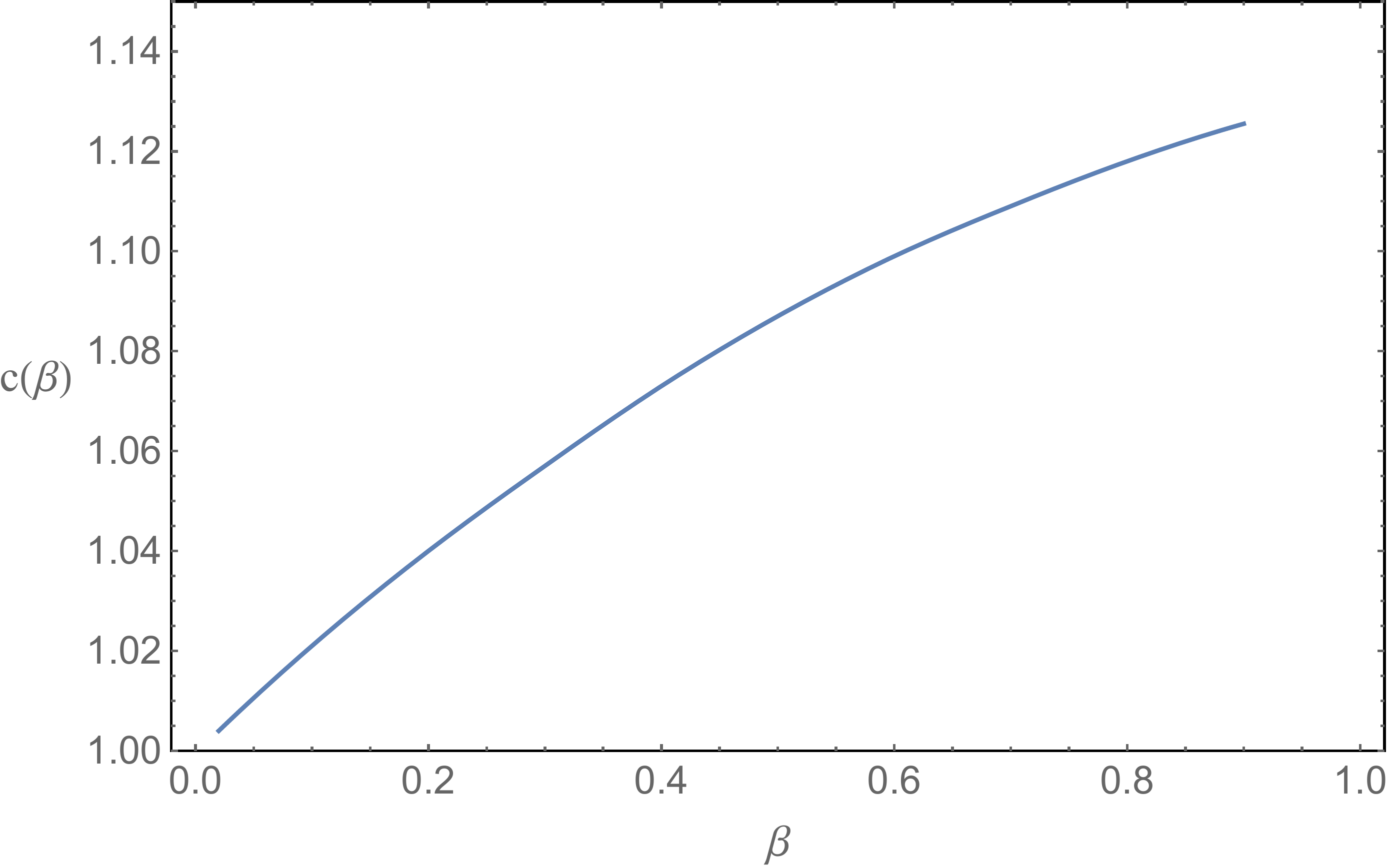}
  \caption{\small $c(\beta)$ defined in (\ref{eq:cbeta}) as a function of the anisotropic scaling exponent $\beta$ in the quantum critical region.}
  \label{fig:c(beta)}
\end{figure}

Due to the extremely small numerical error (\ref{eq:err})
we expect that the value $c(\beta)$ is universal at low temperature and it does not depend on the details of the dual theory while it only depends on the Lifshitz exponent $\beta$ at low energy.
In Fig. \ref{fig:c(beta)} we numerically show the $\beta$-dependence of this coefficient function $c(\beta)$ in a series of strongly coupled Lifshitz-like quantum critical systems.
%It is convenient to introduce the coefficient function $c(\beta)$ such that \begin{equation}\label{eq:def c(beta)} c(\beta)=\frac{\tilde{\sigma}_V}{\tilde{\sigma}_\text{AHE}\hat{T}^2}\bigg{|}_{\hat{T}=0.03}\,.\end{equation}
%As was shown in Fig. \ref{fig:ncveT},
%The value of $c(\beta)$ as a function of $\beta$ is shown in Fig. \ref{fig:c(beta)}.
 $c(\beta)$ monotonically increases with respect to $\beta$.\footnote{The odd (Hall) viscosity has been studied recently in \cite{Copetti:2019rfp} by considering the field theory for critical Lifshitz fermions at finite temperature, which is different from our study where the Lifshitz symmetry is an exact symmetry only at zero temperature in the IR fixed point. In our study there will always be a small deviation from Lifshitz symmetry due to some irrelevant deformations at low and finite temperature.} %Also, it only gives rise to slight variation when $\beta$ changes from $0$ to $1$, i.e., $\frac{\partial f(\beta)}{\partial\beta}\lesssim0.2$. This is an important difference between chiral vortical conductivity and odd viscosities, since similar coefficients of odd viscosities show great changes.
It would be very intriguing to compute $c(\beta)$ from kinetic theory and hydrodynamics to compare to this numerical result from the holographic approach.

%%%%%%%%%%%%%%%%%%%%%%%%
\vspace{.3cm}
%%%%%%%%%%%%%%%%%%%%%%%%

\begin{figure}[h!]
  \centering
\includegraphics[width=0.49\textwidth]{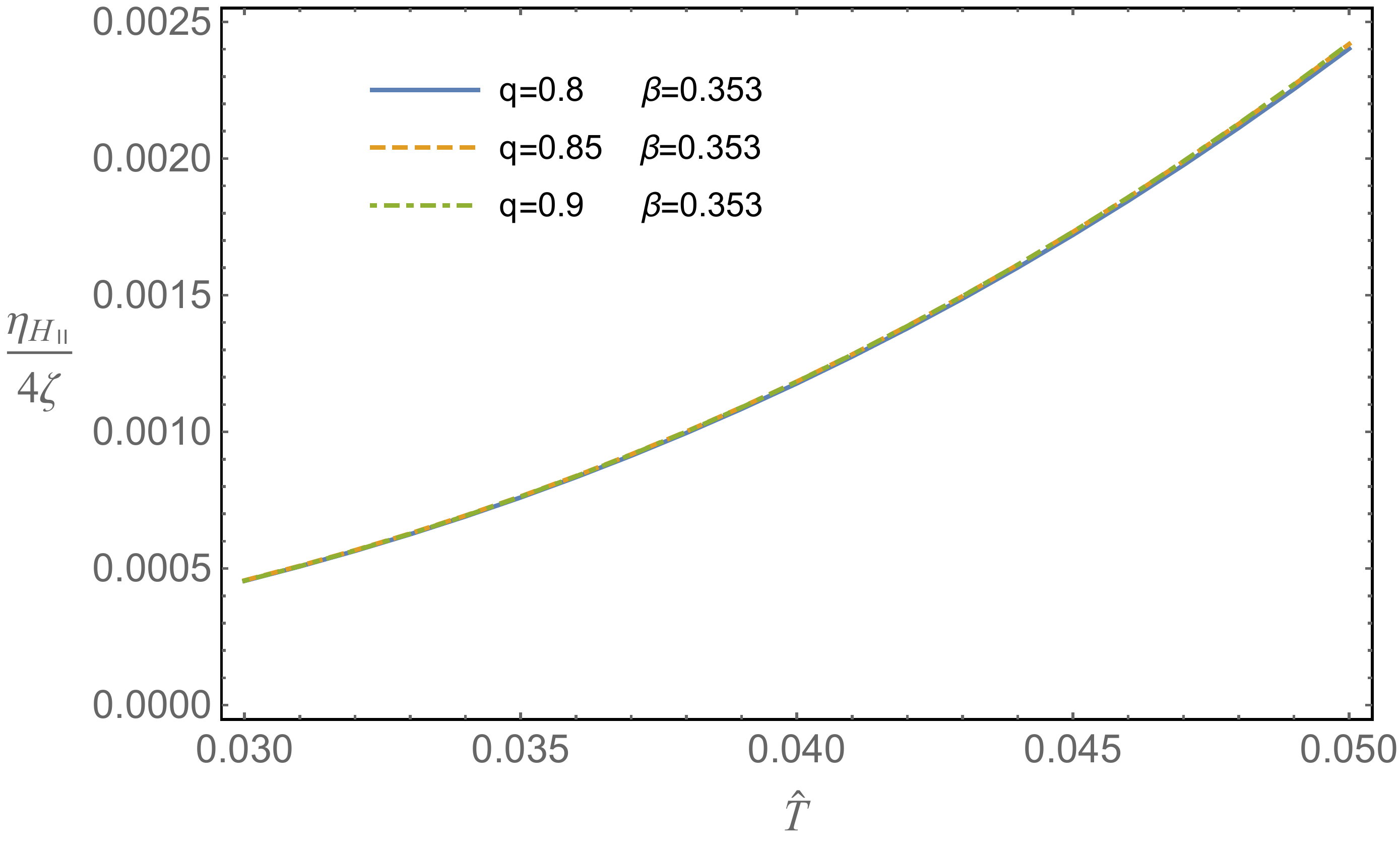}
\includegraphics[width=0.49\textwidth]{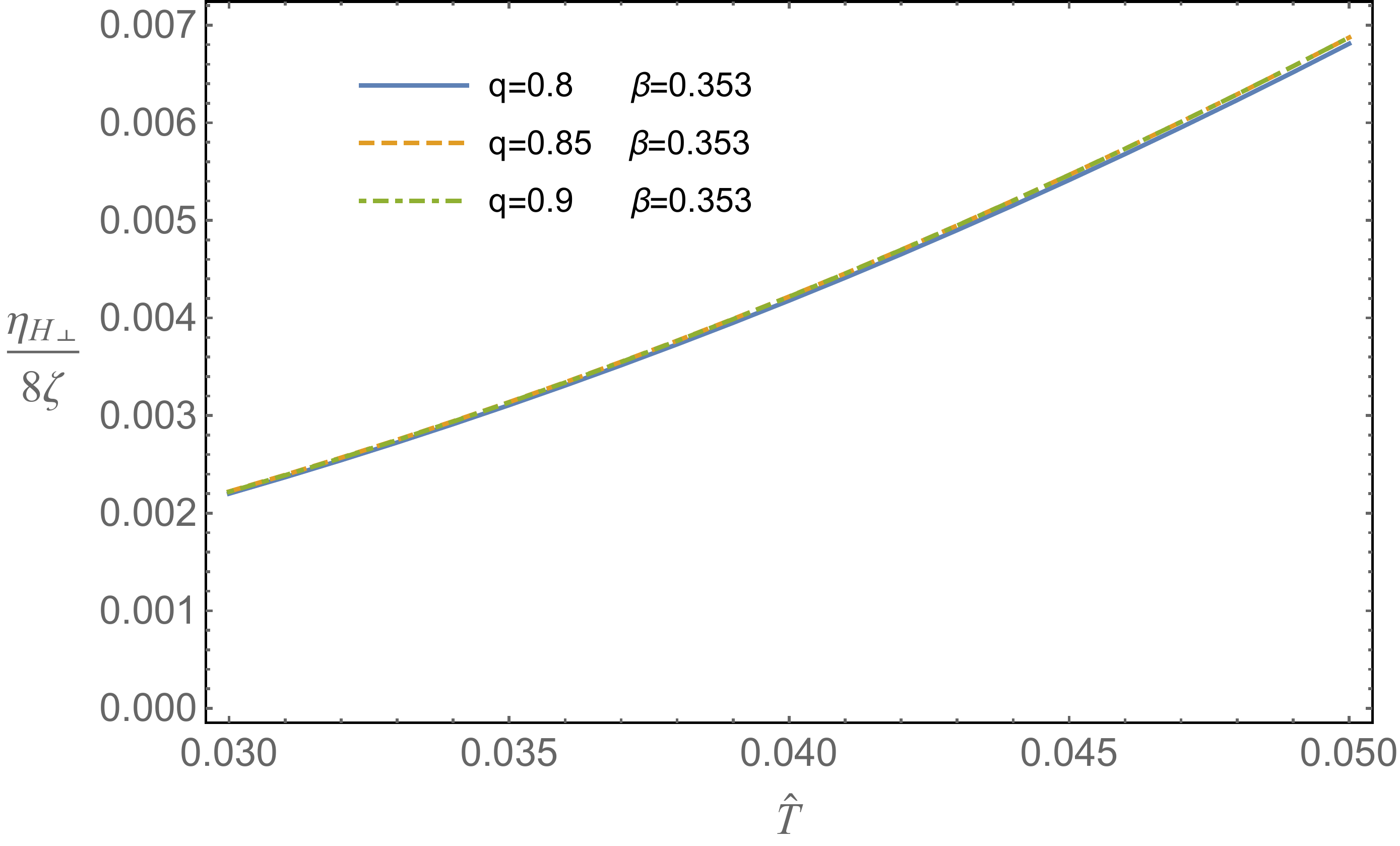}
  \caption{\small Universality of odd viscosities $\eta_{H_\parallel}$ and $\eta_{H_\perp}$ at different QCPs with the same $\beta=0.353$. These two mixed anomaly induced viscosities only depend on the scaling factor $\beta$ in the quantum critical region.}
  \label{fig:hallT}
\end{figure}

It is expected that similar universalities exist for other mixed gauge-gravitational anomaly induced transports such as odd (Hall) viscosities $\eta_{H_\parallel}$ and $\eta_{H_\perp}$ in the quantum critical region, as is shown in Fig. \ref{fig:hallT}. Different from the chiral vortical conductivity, these two odd viscosities can be obtained directly from the near-horizon geometry \cite{Landsteiner:2016stv}. In addition to the universal behaviors relying on the anisotropic scaling exponent $\beta$ as $\hat{T}\rightarrow 0$, one can also notice that as the temperature is increased, these behaviors gradually disappear.
%\red{get lost} \blue{become unobvious}.
This can be viewed as an evidence that the thermal effect depresses the quantum effect. One can again examine the coefficient functions from these two odd viscosities. %\blue{
However, different from the chiral vortical conductivity, the related numerical errors for odd viscosities are much larger in the regime of the temperature we considered, which makes numerical results not very useful. It is expected that when the temperature is sufficiently low the numerical error will be under control and there will be similar universalities in the coefficients functions as the case of chiral vortical conductivity. We will not discuss them here.

%%%%%%%%%%%%%%%%%%%%%%%%
\vspace{.3cm}
%%%%%%%%%%%%%%%%%%%%%%%%

%%%%%%%%%%%%%%%%%%%%%%%%%%
\section{Conclusions}
\label{sec4}
%%%%%%%%%%%%%%%%%%%%%%%%%%
We have studied the chiral vortical conductivity in a holographic Weyl semimetal model \cite{Landsteiner:2015pdh} in which there exists a quantum phase transition between the strongly coupled topologically nontrivial phase and a trivial phase. Motivated by \cite{Copetti:2016ewq} we renormalized the chiral vortical conductivity by the anomalous Hall conductivity and temperature squared, and explored the universal physics of this ratio at low temperature. We found in the Weyl semimetal phase, at low temperature the renormalized chiral vortical conductivity $\tilde{\sigma}_V/(\tilde{\sigma}_\text{AHE}\hat{T}^2)$ stays a constant. In the quantum critical region, the renormalized chiral vortical conductivity is also a constant value, which only depends on the IR scaling exponent $\beta$ at the quantum critical point, while independent of the microscopic details of the dual theory, e.g. we can get the same constant by choosing different values of $\lambda$, $q$ which give rise to different $M_c/b$ but with  the same IR scaling exponent $\beta$. In the topologically trivial phase, for sufficiently large $M/b$ the renormalized chiral vortical conductivity goes to another constant which typically depends on the details of the theory.

The universality in the quantum critical region is quite intriguing. The numerical evidences from holography indicate that the scaling exponent $\beta$ is the only parameter for the mixed gauge-gravitational anomaly induced chiral vortical conductivity discussed in Sec \ref{sec3}. These convincible evidences from holography might motivate one to consider the role of mixed gauge-gravitational anomaly in the Lifshitz-like critical systems for weakly-coupled theories and also from other computational approaches. A probable alternative approach is to study the quantum field theory consisting of a massive fermion coupled to an axial $U(1)$ field  with a background $U(1)_A$ field along spatial directions and calculate the chiral vortical conductivity and the anomalous Hall conductivity at finite temperature to study the universalities of these transports.

%%%%%%%%%%%%%%%%%%%%%%%%%%%%%%%%%%%%%%%%%
\subsection*{Acknowledgments}
%%%%%%%%%%%%%%%%%%%%%%%%%%%%%%%%%%%%%%%%%
We would like to thank Rong-Gen Cai,  Karl Landsteiner, Francisco Pena-Benitez,  and Ya-Wen Sun for useful discussions. This work has been supported by the Thousand Young Talents Program of China. %The work of X.J was also supported by the University of Chinese Academy of Sciences. The work of Y.L. and X.M.W was also supported by a grant from Beihang University and NFSC Grant No. 11875083.
The work of X.J. was also supported by the University of Chinese Academy of Sciences through Grant No. Y95402XXX2. The work of Y.L. and X.M.W. was also supported by a grant from Beihang University and the National Natural Science Foundation of China (NSFC) Grant No. 11875083.

\begin{appendix}

%%%%%%%%%%%%%%%%%%%
\section{Asymptotic expansions for the background fields}
\label{app:a}
%%%%%%%%%%%%%%%%%%%

Near conformal boundary, the background fields can be expanded as
\begin{eqnarray}
&&u= r^{2}-\frac{M^{2}}{3}+\frac{M^{4}\left(2+3\lambda\right)}{18}\frac{\ln r}{r^{2}}-\frac{M_{b}}{3r^{2}}+\dots \nonumber\\
&&f= r^{2}-\frac{M^{2}}{3}+\frac{M^{4}\left(2+3\lambda\right)}{18}\frac{\ln r}{r^{2}}-\frac{f_{3}}{r^{2}}+\dots \nonumber\\
&&h= r^{2}-\frac{M^{2}}{3}+\left(\frac{M^{4}\left(2+3\lambda\right)}{18}+\frac{q^{2}b^{2}M^{2}}{2}\right)\frac{\ln r}{r^{2}}+\frac{h_{3}}{r^{2}}+\dots \nonumber\\
&&A_{z}= b-bM^{2}q^{2}\frac{\ln r}{r^{2}}+\frac{\zeta}{r^{2}}+\dots \nonumber\\
&&\phi= \frac{M}{r}-\frac{\ln r}{6r^{3}}\left(2M^{3}+3b^{2}Mq^{2}+3M^{3}\lambda\right)+\frac{O}{r^{3}}+\dots\,. \nonumber
\end{eqnarray}

%%%%%%%%%%%%%%%%%%%
\section{Calculations of chiral vortical conductivity}
\label{app:cve}
%%%%%%%%%%%%%%%%%%%

We outline the details of calculating the chiral vortical conductivity. Note that in the field theory the chiral vortical conductivity can be obtained by Kubo formulae (\ref{eq:kubofor}).
In holography, we consider the following fluctuations
\be
\delta g_{jz}=\delta g_{zj}=h_{jz}\left(r\right)e^{ik_zz} \,,~~~
\delta g_{tj}=\delta g_{jt}=h_{tj}\left(r\right)e^{ik_zz} \,,~~~
\delta A_{j}=a_{j}\left(r\right)e^{ik_zz} \,,
\ee
with $j\in\{x,y\}.$ The equations of motion are
\bea
h_{ti}^{\prime\prime}+\frac{h^{\prime}}{2h}h_{ti}^{\prime}+C_{1}h_{ti}-\frac{2i\zeta k_{z}}{\sqrt{h}}\epsilon_{ij}\bigg[D_1\left(h_{jz}^{\prime}-\frac{f'}{f}h_{jz}\right)
+E_{1}a_{j}^{\prime}+F_{1}a_{j}\bigg]&=&0\,,\\
h_{iz}^{\prime\prime}+\left(\frac{u^{\prime}}{u}-\frac{h^{\prime}}{2h}\right)h_{iz}^{\prime}+\left(\frac{f^{\prime}h^{\prime}}{2fh}-\frac{f^{\prime\prime}}{f}-\frac{f^{\prime}u^{\prime}}{fu}\right)h_{iz}
+A_{z}^{\prime}a_{i}^{\prime}+\frac{2q^{2}\phi^{2}A_{z}}{u}a_{i}~~~&&\nn\\\label{eq:b3}
-\frac{2i\zeta k_{z}}{\sqrt{h}}\epsilon_{ij}\left[A_{z}^{\prime}\left(\frac{u^{\prime}}{u}-\frac{f^{\prime}}{f}\right)h_{tj}^{\prime}+G_{1}h_{tj}\right]&=&0\,,\\
a_{i}^{\prime\prime}+\left(\frac{u^{\prime}}{u}+\frac{h^{\prime}}{2h}\right)a_{i}^{\prime}-\left(\frac{2q^{2}\phi^{2}}{u}+\frac{k_{z}^{2}}{hu}\right)a_{i}-\frac{A_{z}^{\prime}}{h}
\left(h_{iz}^{\prime}-\frac{f^{\prime}}{f}h_{iz}\right)~~~&&\nn\\
-\frac{2i\zeta k_{z}}{\sqrt{h}}\epsilon_{ij}\left(\frac{E_{1}}{u}h_{tj}^{\prime}+H_{1}h_{tj}\right)&=&0\,,
\end{eqnarray}
and constraint equation
\be\label{eq:constr}
h_{iz}'-\frac{f'}{f}h_{iz}+A_z' a_i-\frac{2i\zeta k_{z}}{\sqrt{h}}
\epsilon_{ij}\left(\frac{u^{\prime}}{u}-\frac{f^{\prime}}{f}\right)A_z'h_{tj}=0\,,
\ee
where $i,j\in\{x,y\}$, $\epsilon_{xy}=-\epsilon_{yx}=1$,
and the coefficients $C_1, D_1, E_1, F_1, G_1, H_1$ are
\begin{eqnarray}
&&C_{1}=-\frac{f^{\prime\prime}}{f}-\frac{f^{\prime}h^{\prime}}{2fh}-\frac{k_{z}^{2}}{hu}\,,\nonumber\\
&&D_1=\frac{uA_{z}^{\prime}}{ h}\left(\frac{u^{\prime}}{u}-\frac{f^{\prime}}{f}\right)\,,\nn\\
&&E_{1}=-u\left(\frac{u^{\prime}}{u}-\frac{f^{\prime}}{f}\right)
\left(\frac{2h^{\prime}}{h}+\frac{f^{\prime}}{f}\right)\,,\nn\\
&&F_{1}=2u^{\prime\prime\prime}-\frac{2uf^{\prime\prime\prime}}{f}-\frac{2uf^{\prime3}}{f^{3}}+u^{\prime}\left(\frac{2f^{\prime2}}{f^{2}}-\frac{4f^{\prime\prime}}{f}\right)+\frac{4uf^{\prime}f^{\prime\prime}}{f^{2}}\,,\nonumber\\
&&G_{1}=\left(\frac{u^{\prime}}{u}-\frac{f^{\prime}}{f}\right)
\left(A_{z}^{\prime\prime}-
\frac{3h^{\prime}}{2h}A_{z}^{\prime}\right)\,,\nonumber\\
%&&H_{1}=\frac{f^{\prime}}{f}\left(\frac{f^{\prime}}{f}+\frac{2h^{\prime}}{h}-\frac{u^{\prime}}{u}\right)-\frac{2h^{\prime}u^{\prime}}{hu}\nonumber\\
&&H_{1}=\left(\frac{u^{\prime}}{u}-\frac{f^{\prime}}{f}\right)
\left[\left(\frac{f^{\prime\prime}}{f}-\frac{h^{\prime\prime}}{h}\right)+
\frac{3h^{\prime}}{2h}\left(\frac{f^{\prime}}{f}+\frac{h^{\prime}}{h}\right)\right]\,.\nn
\end{eqnarray}
It is easy to show that (\ref{eq:constr}) is consistent with (\ref{eq:b3}) by using the equation of motion for the background fields. Thus $h_{iz}$ is completely determined by $a_i$ and $h_{tj}$.

The above equations can be further simplified as
\bea
h_{ti}^{\prime\prime}+\frac{h^{\prime}}{2h}h_{ti}^{\prime}+\left(C_{1}-\frac{2\zeta^2 k_z^2}{u}D_1^2 \right)h_{ti}-\frac{2i\zeta k_{z}}{\sqrt{h}}\epsilon_{ij}\bigg[
E_{1}a_{j}^{\prime}+\left(F_{1}-D_1 A_z'\right) a_j\bigg]
&=&0\,,\nn\\
%~~&&\nn\\
%-\frac{2\zeta^2 k_z^2}{h}D_1\left(\frac{u'}{u}-\frac{f'}{f}\right)A_z' h_{ti}&=&0\,,\\
a_{i}^{\prime\prime}+\left(\frac{u^{\prime}}{u}+\frac{h^{\prime}}{2h}\right)a_{i}^{\prime}-\left(\frac{2q^{2}\phi^{2}}{u}+\frac{k_{z}^{2}}{hu}-\frac{A_z'^{2}}{h}\right)a_{i}
%-\frac{A_{z}^{\prime}}{h}\left(h_{iz}^{\prime}-\frac{f^{\prime}}{f}h_{iz}\right)
%~~~&&\nn\\
-\frac{2i\zeta k_{z}}{\sqrt{h}}\epsilon_{ij}\left[\frac{E_{1}}{u}h_{tj}^{\prime}+\left(H_{1}+\frac{D_1}{u}A_z'\right)h_{tj}\right]&=&0\,.\nn
\eea

One can solve them order by order
\be
h_{ti}=h_{ti}^{(0)}+k_z h_{ti}^{(1)}+k_z^2h_{ti}^{(2)}+\dots\,,~~~
a_{i}=a_{i}^{(0)}+k_z a_{i}^{(1)}+k_z^2 a_{i}^{(2)}+\dots\,.~~~
\ee
We consider the sector $h_{ti}^{(0)}=u, a_i^{(0)}=0$,
thus $h_{ti}^{(1)}=0$, and the equation for $a_i^{(1)}$ is
\bea\label{eq:appfluc}
a_{i}^{(1)\prime\prime}+\left(\frac{u^{\prime}}{u}+\frac{h^{\prime}}{2h}\right)a_{i}^{(1)\prime}-\left(\frac{2q^{2}\phi^{2}}{u}-\frac{A_z'^{2}}{h}\right)a_{i}^{(1)}
-\frac{2i\zeta }{\sqrt{h}}\epsilon_{ij}\left(\frac{u'}{u}-\frac{f'}{f}\right)L_1&=&0
\eea
with $$L_1=\left[-\left(\frac{2h'}{h}+\frac{f'}{f}\right)h_{tj}^{(0)\prime}+\left(\left(\frac{f^{\prime\prime}}{f}-\frac{h^{\prime\prime}}{h}\right)+
\frac{3h^{\prime}}{2h}\left(\frac{f^{\prime}}{f}+\frac{h^{\prime}}{h}\right)+\frac{A_z'^2}{h}\right)h_{tj}^{(0)}\right]\,.$$
\subsection{Asymptotic expansion}

Near boundary, the fluctuations can be expanded as
\begin{eqnarray}
h_{ti}&=&h_{ti}^{\left[0\right]}r^{2}-h_{ti}^{\left[0\right]}\left(\frac{k_z^{2}}{4}+\frac{M^{2}}{3}\right)+\frac{\ln r}{r^{2}}\left[\frac{h_{ti}^{\left[0\right]}}{144}\left(9k_z^{4}+16M^{4}+24M^{4}\lambda+6M^{2}k_z^{2}\right)\right]+\frac{h_{ti}^{\left[2\right]}}{4r^{2}}+\dots \nonumber\\
h_{iz}&=&h_{iz}^{\left[0\right]}r^{2}-h_{iz}^{\left[0\right]}\frac{M^{2}}{3}+\frac{\ln r}{r^{2}}\left[\frac{h_{iz}^{\left[0\right]}}{18}\left(2M^{4}+3M^{4}\lambda\right)+2a_{i}^{\left[0\right]}bM^{2}q^{2}\right]+\frac{h_{iz}^{\left[2\right]}}{4r^{2}}+\dots \nonumber\\
a_{i}&=&a_{i}^{\left[0\right]}-a_{i}^{\left[0\right]}\left(M^{2}q^{2}+\frac{k_z^{2}}{2}\right)\frac{\ln r}{r^{2}}+\frac{a_{i}^{\left[2\right]}}{r^{2}}+\dots \nonumber\\
v_{i}&=&v_{i}^{\left[0\right]}-\frac{k_z^{2}}{2}v_{i}^{\left[0\right]}\frac{\ln r}{r^{2}}+\frac{v_{i}^{\left[2\right]}}{r^{2}}+\dots
\end{eqnarray}
where $i\in\{x,y\}$.

\subsection{On shell action}
The renormalised action is given by $S_\text{ren}=S+S_\text{bnd}$ with
\bea
S _{\text{bnd}}&=&\frac{1}{\kappa^2}\int_{r=r_\infty} d^4x \sqrt{-\gamma} K-\frac{1}{2\kappa^2}\int_{r=r_\infty} d^4 x\sqrt{-\gamma}\bigg[6+\frac{1}{2}R+...\bigg]
\nn\\&&~~~
+\frac{\log r}{4}\int_{r=r_\infty} d^4x \sqrt{-\gamma}\bigg[\mathcal{F}_{\mu\nu}\mathcal{F}^{\mu\nu}+F_{\mu\nu}F^{\mu\nu}+|D_m\Phi|^2+\bigg(\frac{1}{3}+\frac{\lambda}{2}\bigg)|\Phi |^4\bigg]
\nn
\eea
where $\gamma_{ab}$ is the induced metric on the boundary, $K$ and $R$ are extrinsic and intrinsic curvatures respectively.
The renormalised on-shell action contains the following terms at quadratic order in fluctuations
\begin{align}\
S_{\text{on-shell}} \supset \int\frac{dk_z}{2\pi}d^3x&\bigg[2a_x^{[0]}(-k_z)a_x^{[2]}(k_z)+2a_y^{[0]}(-k_z)a_y^{[2]}(k_z)\nonumber\\
&-h_{tx}^{[0]}(-k_z)h_{tx}^{[2]}(k_z)-h_{ty}^{[0]}(-k_z)h_{ty}^{[2]}(k_z)\nonumber\\
&+h_{xz}^{[0]}(-k_z)h_{xz}^{[2]}(k_z)+h_{yz}^{[0]}(-k_z)h_{yz}^{[2]}(k_z)\nonumber\\
&+\mathcal{O}(k_z^2)+\text{contact~terms}\bigg]\,,
\end{align}
where
\begin{align}\
\text{contact~terms}=&\left(h_{tx}^{[0]}(-k_z)h_{tx}^{[0]}(k_z)+h_{ty}^{[0]}(-k_z)h_{ty}^{[0]}(k_z)\right)
\left(4f_3-\frac{7M^4}{36}+2MO-M_b\right)\nonumber\\
&+\left(h_{xz}^{[0]}(-k_z)h_{xz}^{[0]}(k_z)+h_{yz}^{[0]}(-k_z)h_{yz}^{[0]}(k_z)\right)
\left(4f_3-\frac{7M^4}{12}+2MO-\frac{M_b}{3}-\frac{M^4\lambda}{2}\right)\nonumber\\
&+M^2q^2\left(a_x^{[0]}(-k_z)a_x^{[0]}(k_z)+a_y^{[0]}(-k_z)a_y^{[0]}(k_z)\right)\nonumber\\
&+\left(a_x^{[0]}(-k_z)h_{xz}^{[0]}(k_z)+a_y^{[0]}(-k_z)h_{yz}^{[0]}(k_z)\right)\left(-bM^2q^2-2\xi\right)\nonumber\\
&-\frac{1}{2}bM^2q^2\left(h_{xz}^{[0]}(-k_z)a_x^{[0]}(k_z)+h_{yz}^{[0]}(-k_z)a_y^{[0]}(k_z)\right)+\mathcal{O}(k_z^2)\,.
\end{align}

With the sourceless boundary condition for $a_i$, i.e. $a_i^{[0]}=0$, we normalized the source of $h_{tj}$ to be $h_{tj}^{[0]}=1$,
where $i,j\in \{x, y\}$ and $i\neq j$
we have
\be G_{i,tj}^{R}=2a_i^{[2]}\,.\ee

%It is obvious that CVE comes from the first two terms from which we can define the corresponding retarded Green's function,
%\begin{equation}
%2a_x^{(0)}(-k_z)a_x^{(2)}(k_z)=a_x^{(0)}(-k_z)G_{ty,x}^{R}h_{ty}^{(0)}(k_z),
%\end{equation}
%\begin{equation}
%2a_y^{(0)}(-k_z)a_y^{(2)}(k_z)=a_y^{(0)}(-k_z)G_{tx,y}^{R}h_{tx}^{(0)}(k_z),
%\end{equation}
%Thus if we normalize the source term $h_{ty}^{(0)}$ to be $1$, then we have
%\begin{equation}\label{Green's function}
%G_{ty,x}^{R}=2a_x^{(2)},~~~G_{tx,y}^{R}=2a_y^{(2)}.
%\end{equation}

\end{appendix}
\appendix

\end{document}